\documentstyle[12pt]{article}
\def\be{\begin{equation}}
\def\ee{\end{equation}}
\def\bc{\begin{center}}
\def\ec{\end{center}}
\def\bea{\begin{eqnarray}}
\def\eea{\end{eqnarray}}
\def\atu{{\alpha_t^U}}
\def\atauu{{\alpha_{\tau}^U}}
\def\abu{{\alpha_b^U}}
\def\at{{\alpha_t}}
\def\ab{{\alpha_b}}
\def\atau{{\alpha_{\tau}}}
\def\htu{{h_t^U}}
\def\htauu{{h_{\tau}^U}}
\def\hbu{{h_b^U}}

\def\gev{{\rm \; GeV}}

\def\mpl{M_{\rm Planck}}
\def\msu{{M_{\rm SUSY}}}
\def\simlt{\stackrel{<}{{}_\sim}}
\def\simgt{\stackrel{>}{{}_\sim}}
\def\tb{{\tan \beta}}

\catcode`@=11
\def\marginnote#1{}
\newcount\hour
\newcount\minute
\newtoks\amorpm
\hour=\time\divide\hour by60
\minute=\time{\multiply\hour by60 \global\advance\minute by-\hour}
\edef\standardtime{{\ifnum\hour<12 \global\amorpm={am}%
        \else\global\amorpm={pm}\advance\hour by-12 \fi
        \ifnum\hour=0 \hour=12 \fi
        \number\hour:\ifnum\minute<10 0\fi\number\minute\the\amorpm}}
\edef\militarytime{\number\hour:\ifnum\minute<10 0\fi\number\minute}
\def\draftlabel#1{{\@bsphack\if@filesw {\let\thepage\relax
   \xdef\@gtempa{\write\@auxout{\string
      \newlabel{#1}{{\@currentlabel}{\thepage}}}}}\@gtempa
   \if@nobreak \ifvmode\nobreak\fi\fi\fi\@esphack}
        \gdef\@eqnlabel{#1}}
\def\@eqnlabel{}
\def\@vacuum{}
\def\draftmarginnote#1{\marginpar{\raggedright\scriptsize\tt#1}}
\def\draft{\oddsidemargin 0.0truein
        \def\@oddfoot{\sl preliminary draft \hfil
        \rm\thepage\hfil\sl\today\quad\militarytime}
        \let\@evenfoot\@oddfoot \overfullrule 3pt
        \let\label=\draftlabel
        \let\marginnote=\draftmarginnote
   \def\@eqnnum{(\theequation)\rlap{\kern\marginparsep\tt\@eqnlabel}%
\global\let\@eqnlabel\@vacuum}  }
\catcode`@=12
\begin{document}
\begin{titlepage}
\vspace*{-1cm}
\phantom{bla}
\hfill{CERN-TH/95-11}
\\
\phantom{bla}
\hfill{LPTENS-95/06}
\\
\phantom{bla}
\hfill{hep-ph/9502318}
\vskip 1.5cm
\begin{center}
{\Large\bf Possible dynamical determination of $m_t$, $m_b$ and
$m_{\tau}$}
\footnote{Work supported in part by the European Union under
contracts No.~CHRX-CT92-0004, SC1$^*$-0394C and SC1$^*$-CT92-0789.}
\end{center}
\vskip 1.0cm
\begin{center}
{\large Costas Kounnas}\footnote{On leave from the Laboratoire de
Physique Th\'eorique, ENS, Paris, France.}
\\
\vskip .1cm
Theory Division, CERN, CH-1211 Geneva 23, Switzerland
\\
\vskip .2cm
{\large Ilarion Pavel}
\\
\vskip .1cm
Laboratoire de Physique Th\'eorique, ENS, F-75231 Paris Cedex 05,
France
\\
\vskip .2cm
{\large Giovanni Ridolfi}\footnote{On leave from INFN, Sezione di
Genova, Genoa, Italy.}
\\
\vskip .1cm
Theory Division, CERN, CH-1211 Geneva 23, Switzerland
\\
\vskip .2cm
and
\\
\vskip .2cm
{\large Fabio
Zwirner}\footnote{On leave from INFN, Sezione di Padova, Padua,
Italy.}
\\
\vskip .1cm
Theory Division, CERN, CH-1211 Geneva 23, Switzerland
\end{center}
\vskip 0.5cm
\begin{abstract}
\noindent
Motivated by four-dimensional superstring models, we consider the
possibility of treating the Yukawa couplings of the Minimal
Supersymmetric Standard Model (MSSM) as dynamical variables of the
effective theory at the electroweak scale. Assuming bottom-tau
unification, we concentrate on the top and bottom Yukawa couplings,
and find that minimizing the effective potential drives them close to
an effective infrared fixed line. Requiring an acceptable bottom-top
mass ratio leads in principle to an additional constraint on the
MSSM parameter space. As a by-product, we give new approximate
analytical solutions of the renormalization group equations for
the MSSM parameters.
\end{abstract}
\vfill{
CERN-TH/95-11
\newline
\noindent
January 1995}
\end{titlepage}
\setcounter{footnote}{0}
\vskip2truecm
\vspace{1cm}
{\bf 1.}
In a recent paper [\ref{kpz}], the possibility was discussed
of treating the Yukawa couplings of the Minimal Supersymmetric
Standard Model (MSSM) not as numerical parameters but as
dynamical variables (for similar suggestions, see also
[\ref{similar},\ref{binetruy}]).
This possibility naturally arises when one embeds the MSSM into
a more fundamental theory, such as supergravity or superstrings,
where parameters are replaced by vacuum expectation values of some
singlet scalar fields (moduli), corresponding to approximately flat
directions of the effective potential.
In [\ref{kpz}], the discussion was mainly restricted to the
top-quark Yukawa coupling, and it was found that, if the scale
$\msu$ of the explicit MSSM mass terms is of the order of the
electroweak scale, and if supersymmetry breaking does not induce
mass terms larger than ${\cal O} (\msu^2/\mpl)$ in the relevant
moduli directions, then minimizing the vacuum energy attracts the
top-quark Yukawa coupling close to its effective infrared fixed
point, which is compatible with a top-quark mass in the
experimentally allowed range. In [\ref{kpz}], it was assumed
that the scale $\msu$, proportional to the gravitino mass $m_{3/2}$,
also corresponds to an approximately flat direction of the
fundamental theory, as suggested by a certain class of
supergravity models [\ref{noscale},\ref{lhc}]. However, we
shall see that the result on the top-quark Yukawa
coupling remains valid even if one assumes (as is often
done in phenomenological analyses) that $\msu$ is fixed by
some physics at very high scales, which allows $\msu$
to be treated as an input parameter, independent of the Yukawa
couplings apart from the usual renormalization effects,
in the effective field theory at the electroweak scale.

In the present paper, we examine the possibility of dynamically
explaining the third-generation fermion masses: $m_t$, $m_b$,
$m_{\tau}$. To do so, we generalize the considerations of
[\ref{kpz}] to the case where all the third-generation Yukawa
couplings  are included, still neglecting the two light generations.
Assuming for simplicity the unification relation $h_b (M_U) =
h_{\tau} (M_U)$, which in first approximation fits the experimental
value of the $m_b/m_{\tau}$ ratio [\ref{btauns},\ref{btaususy}],
and treating the explicit MSSM mass terms as numerical parameters,
we find that in this more complicated case the Yukawa couplings
are dynamically attracted close to an effective infrared fixed
line, $F(h_t,h_b)=0$. We also find that minimization of the
effective potential with respect to the residual variable $\theta$,
which parametrizes the infrared fixed line, dynamically fixes the
ratio $h_b/h_t$, and may allow for acceptable values of the $m_b/
m_t$ ratio within the residual MSSM parameter space. Our approach
leads to the elimination of two of the free parameters of the
MSSM, $h_t$ and $h_b$, in addition to the parameter $h_{\tau}$,
removed by the unification relation $h_b (M_U) = h_{\tau} (M_U)$.
Furthermore, the infrared behaviour of the renormalization group
equations (RGEs) for the mass parameters of the MSSM is such that
the latter are severely constrained, with some combinations being
driven close to effective infrared fixed values.

The structure of our paper is as follows. In section 2, we
review the infrared behaviour of the running Yukawa couplings
and masses in the MSSM [\ref{hill}--\ref{fl}], presenting some
approximate analytical solutions of the RGE for the top and bottom
Yukawa couplings\footnote{Similar results have been simultaneously
and independently obtained in [\ref{fl}].}, as well as for the
MSSM mass terms, when both $h_b$ and $h_t$ are non-negligible.
In section~3, we present the theoretical motivations that lead us
to minimize the effective potential not only with respect to
the Higgs fields, but also with respect to the top and bottom
Yukawa couplings, taken as independent dynamical variables.
In section~4 we describe in some detail the results of such a
minimization, both analytically and numerically, working for given
numerical boundary conditions on the MSSM mass parameters at the
unification scale $M_U$. We begin by showing that the leading
dependence of the effective MSSM potential on $h_t$ and $h_b$ forces
the latter, upon minimization, to lie close to the effective infrared
fixed line. We then proceed to the more subtle problem of minimizing
the low-energy  effective potential along the infrared fixed line.
In section~5 we summarize our conclusions, after commenting on the
case in which $\msu$ is also taken as a dynamical variable and on the
case in which the top and bottom Yukawa are constrained dynamical
variables.

\vspace{1cm}
{\bf 2.}
Neglecting as announced the first two generations, and working as
usual in a mass-independent renormalization scheme, the one-loop
RGEs for the Yukawa couplings read~[\ref{rge}]
\be
\label{rgeyuk}
\begin{array}{ccl}
\displaystyle{
\frac{d \alpha_t}{d t} } & = &  \displaystyle{ \frac{\alpha_t}{ 4\pi}
\left(
\frac{16}{3} \alpha_3
+ 3 \alpha_2
+ \frac{13}{9} \alpha'
- 6 \alpha_t
- \alpha_b
\right) \, , }
\\ & & \\
\displaystyle{
\frac{d \alpha_b}{d t} } & = & \displaystyle{
\frac{\alpha_b}{4 \pi}
\left(
\frac{16}{3} \alpha_3
+ 3 \alpha_2
+ \frac{7}{9} \alpha'
- \alpha_t
- 6 \alpha_b
- \alpha_{\tau}
\right) \, ,}
\\ & & \\
\displaystyle{
\frac{d \alpha_{\tau}}{d t} }
& = & \displaystyle{ \frac{\alpha_{\tau}}{4 \pi}
\left(
3 \alpha_2
+ 3 \alpha'
- 3 \alpha_b
-  4 \alpha_{\tau}
\right) \, , }
\end{array}
\ee
where $\alpha_{t,b,\tau} \equiv h_{t,b,\tau}^2 / (4 \pi)$, $t
\equiv \log(M_U^2/Q^2)$ and $M_U \simeq 2 \times 10^{16} \gev$.
Two-loop RGEs are available [\ref{twoloop}], but we do not need
them for our present purpose.

If in eqs.~(\ref{rgeyuk}) we neglect $\ab$ and $\atau$ with respect
to $\at$, which is a good first approximation when $\tb \equiv v_2/
v_1 \ll m_t/m_b$, and take $Q \sim \msu \sim m_Z$, then the effective
infrared fixed point for the top-quark Yukawa coupling is
[\ref{apw},\ref{numacc}] $\at \simeq (8/9) \alpha_3$, corresponding
to a running top-quark mass $m_t \simeq (195 \gev) \sin \beta$, from
which the pole mass can be extracted by including the standard QCD
corrections, $m_t^{pole} = m_t(m_t) [ 1 + 4 \alpha_3(m_t) / (3 \pi)
+ \ldots ]$. When $\ab$ and $\atau$ are not neglected, which is the
case of interest for the present paper, eqs.~(\ref{rgeyuk}) have a
more complicated infrared behaviour. Choosing for simplicity $\abu
= \atauu$, in order to have a more manageable two-variable problem,
the resulting infrared structure is displayed in fig.~1. We choose
random boundary conditions satisfying the constraint
\be
2 \alpha_U < \atu + \abu < 1 \, ,
\ee
with $\alpha_U \simeq 1/25$, corresponding to the dots in the region
of the $(\atu,\abu)$ plane shown in fig.~1a. We then solve
numerically the RGEs of eq.~(\ref{rgeyuk}) at the representative
scale $Q = 200 \gev$, to obtain $\at$, $\ab$ and $\atau$. The
resulting region of the $(h_t,h_b)$ plane is shown by the dots in
fig.~1b. The important aspect to be stressed is the focusing effect
due to the infrared structure of the RGE: a relatively wide region of
the $(\htu,\hbu)$ plane is mapped into a very thin region of the
$(h_t,h_b)$ plane, clustering around an `effective infrared fixed
line'. Another effect, clearly visible in fig.~1, is the existence
of some special points along the effective infrared fixed line. If
we look at the density of points in the $(h_t,h_b)$ plane,
corresponding to a uniform distribution in the $(\htu,\hbu)$ plane,
we can clearly see that the point $h_t=h_b$ is an attractor, whereas
$h_t=0$ and $h_b=0$ are repulsors.

For practical purposes, we now introduce some approximate analytical
formulae for $h_t$ and $h_b$, which can be useful to parametrize the
effective infrared fixed line. If in the RGEs for $\at$ and $\ab$ we
neglect the terms proportional to $\atau$ and to $\alpha'$, and we
define the auxiliary variables
\be
\label{rhotheta}
\rho \equiv \sqrt{h_t^2 + h_b^2} \, ,
\;\;\;\;\;\;\;
\tan \theta \equiv {h_b \over h_t} \, ,
\ee
after some calculations we can write the solution as
\begin{eqnarray}
\label{part1}
{f ( \sin^2 2 \theta ) \over \rho^2}
& = &
{f ( \sin^2 2 \theta_U ) \over \rho_U^2} {1 \over E}
+
{3 \over 8 \pi^2} {F \over E} \, , \\
& & \nonumber \\
\label{part2}
{\rho^2 ( \sin 2 \theta )^{12/5} \over (\cos 2 \theta)^{7/5}}
& = &
{\rho_U^2 ( \sin 2 \theta_U )^{12/5} \over (\cos 2 \theta_U)^{7/5}}
\cdot E \, ,
\end{eqnarray}
where
\be
f(x) \equiv \;_2 {\cal F}_1 \left( 1/2,1,11/5 \, ;x \right)
= {6 \over 5} \int_0^1 ds (1-s)^{1/5} (1 - sx)^{-1/2} \, ,
\ee
\be
E (t) \equiv
\left( 1 - {3 \alpha_U \over 4 \pi} t \right)^{-16/9}
\left( 1 + {\alpha_U \over 4 \pi} t \right)^{3} \, ,
\;\;\;\;\;
F \equiv \int_0^{t} dt' E(t') \, .
\ee
The same result was independently obtained in ref.~[\ref{fl}].
Since the behaviour of the hypergeometric function $f$ will play
an important role in the considerations of section~4, for
illustration we plot in fig.~2 $f(\sin^2 2 \theta)$ and $df(\sin^2
2 \theta)/d \theta$, as functions of $\theta$. From
eq.~(\ref{part2}), we can see that the renormalization of the
quantity in the first member is multiplicative. On the other hand,
$1 \le f(\sin^2 2 \theta) \le 12/7$, and at scales $Q  \sim 200
\gev$ it is (still neglecting $\alpha'$ effects) $E \simeq 10$
and $F \simeq 220$, so that the second member of eq.~(\ref{part1})
is dominated by $3F/(8 \pi^2 E)$ for sufficiently large values of
$\rho_U$ ($\rho_U \simgt 0.5$). This defines an effective infrared
fixed line at any scale $Q$ close to the electroweak scale,
parametrized by
\be
\label{hthb0}
h_t = \rho_{IR}(\theta) \cos \theta \, ,
\;\;\;\;\;
h_b = \rho_{IR}(\theta) \sin \theta \, ,
\ee
where
\be
\label{rhoir}
\rho_{IR}(\theta) \equiv
\sqrt{{8 \pi^2 E \over 3 F} f ( \sin^2 2 \theta )}
\, .
\ee
To further improve the approximation of eq.~(\ref{hthb0}), we can
introduce some constant shifts to  fit the corrections due to
$\alpha'$ and $\atau$ effects, and get
\be
\label{hthb}
h_t = 0.015 + \rho_{IR}(\theta) \cos \theta \, ,
\;\;\;\;\;
h_b = - 0.045 + \rho_{IR}(\theta) \sin \theta \, ,
\ee
where of course the  range of variation of $\theta$ should be
modified accordingly. It  is useful to compare our formula with
the one previously derived in [\ref{th7320}], which, after
correcting for $\alpha'$ and $\alpha_{\tau}$ effects as in
eq.~(\ref{hthb}), reads
\be
\label{power}
\left( h_t - 0.015 \right)^{12}
+
\left( h_b + 0.045 \right)^{12}
=
\left( {8 \pi^2 E \over 3 F} \right)^6 \, .
\ee
Figure~1b compares the exact numerical solutions of
eqs.~(\ref{rgeyuk}),
represented by the dots, with the approximate analytical solutions of
eqs.~(\ref{hthb}) and (\ref{power}), represented by the solid and by
the dashed line, respectively. We can see that both formulae are good
approximations for $h_b \ll h_t$ or $h_t \ll h_b$, whereas
eq.~(\ref{hthb}) is a better approximation for $h_b \sim h_t$.

An essential ingredient in the study of the MSSM effective potential
is the solutions to the RGE for the MSSM mass parameters
[\ref{massrge}]. Exact analytical solutions of the one-loop RGE are
known [\ref{extonly}] in the case of negligible $h_b$ and $h_{\tau}$.
Some approximate analytical solutions have also been obtained
recently for the special case $h_b=h_t$ [\ref{largetb}].
We have improved the existing formulae by constructing approximate
analytical solutions valid for any value of $h_t$ and $h_b$, and
including the most important $\alpha'$ effects, but still neglecting
$\alpha_{\tau}$ effects. Their explicit form is given in the
Appendix. From our formulae one can easily rederive some known
relations valid at special points on the effective infrared fixed
curve: $A_t/m_{1/2} \simeq H_8 - H_4/2 \simeq 1.5$ and $\Delta
m_2^2 \simeq (3/2) m_0^2 + (1/2)(H_2 - H_4^2/4) m_{1/2}^2 \simeq
(3/2) m_0^2 + (1/2) (6.3) m_{1/2}^2$ for $\theta = 0$ [\ref{cw}],
$A_t/m_{1/2} \simeq A_b/m_{1/2} \simeq 1.5$ and $\Delta m_1^2
\simeq \Delta m_2^2 \simeq (9/7) m_0^2 + (3/7) (6.3) m_{1/2}^2$ for
$\theta=\pi/4$ [\ref{largetb}].

\vspace{1cm}
{\bf 3.}
We now present the theoretical motivations that lead us to
minimizing the MSSM effective potential not only with
respect to the Higgs fields, but also with respect to the top and
bottom Yukawa couplings, taken as independent dynamical variables.

In a generic supergravity model, masses and couplings are
field-dependent functions. This should be kept in mind when
considering both the dimensionless and the dimensionful
parameters of the MSSM, seen as the low-energy effective theory
of an underlying supergravity model. For each given MSSM parameter,
if the scalar fields that control it are frozen to their VEVs by
sufficiently heavy mass terms, then such a parameter can be treated
as constant, apart from standard renormalization effects, when
discussing the dynamics at the electroweak scale. If, however,
after integrating out the superheavy degrees of freedom, some
extra singlet scalar fields are left, with no renormalizable
couplings to the MSSM fields and masses ${\cal O} (\msu^2 / \mpl)$
or smaller, then MSSM quantum corrections can play a role in the
determination of their VEVs, and the corresponding MSSM parameters
should be treated as dynamical variables of the effective theory
at the electroweak scale.

Consider for example the class of supergravity models whose field
content splits into an `observable sector', containing the MSSM
states (and possibly others) and a `hidden sector', coupled to
the observable sector only via interactions of gravitational
strength. The situation of interest to us can be realized in the
special subclass of models [\ref{noscale},\ref{lhc}] that exhibit,
in their hidden sector, some approximately flat directions of the
classical potential, associated to some `moduli' fields. Such
degeneracy of the classical vacuum is in general removed by
quantum corrections, including the perturbative ones if
supersymmetry is spontaneously broken. We would like to envisage
here the possibility that the potential along some of these flat
directions does not get large quantum corrections from the
superheavy sectors of the theory. Then we need to minimize the
effective potential at the electroweak scale to fix some moduli VEVs
and to determine those MSSM low-energy parameters which carry a
non-trivial dependence on such moduli.

The above possibility is supported by the general structure of
four-dimensional superstring models [\ref{fds}], where all the
low-energy parameters, in particular the Yukawa couplings, are
dynamical variables depending on some moduli VEVs. Indeed, some
interesting superstring solutions could give rise, in the
low-energy limit, to spontaneously broken $N=1$ effective
supergravities of the type considered above. At the classical
level, gauge and Yukawa couplings are related [\ref{yukstr}] by
a string super-unification condition:
\be
\label{su}
\frac{k_i}{\alpha_i}=\frac{1}{\alpha_{str}} \, .
\ee
In eq.~({\ref{su}), $\alpha_{str}$ is the coupling constant
associated with the string loop expansion: already at the
classical level, this is not a numerical parameter but a
dynamical variable, related to the VEV of the dilaton field
$(S + \bar S)$ by $2 \pi \alpha_{str} = (S + \bar S)^{-1}$.
For string solutions with unbroken supersymmetry, ${\alpha_{str}}$
is a flat direction, not only classically but at all orders in the
string perturbative expansion. In the case of the gauge couplings
($i=1,2,3$ for the factors of the standard model gauge group), the
coefficients $k_i$ are constants depending on the particular string
solution. In the following, we shall have in mind the class of string
solutions for which, with the usual normalization convention $g_1 =
\sqrt{5/3} g'$, it is $k_3 = k_2 = k_1 = 1$. We shall also assume
that some non-perturbative effects break spontaneously $N=1$
supersymmetry and fix the VEV of the dilaton field, in such a way
that gauge couplings are not dynamical variables at the electroweak
scale. In the case of the Yukawa couplings ($i=t,b,\tau$ for the
third-generation ones to be considered here), at the  string
classical level the coefficients $k_i$ typically are exponentially
suppressed or of order unity, as can be easily checked in many
explicit examples. For instance, in free fermionic constructions
the non-vanishing tree-level Yukawa couplings correspond to
moduli-independent coefficients $k_i=2$. In other classes of string
solutions one can have moduli-dependent $k_i$ coefficients for the
tree-level Yukawa couplings.

Even when the tree-level moduli-dependence of the Yukawa couplings
is identical to that of the gauge couplings, string-loop
corrections [\ref{thr}] to the low-energy effective action can
introduce additional, non-universal moduli dependences. One can
then envisage, as already discussed in [\ref{kpz}], various
possible situations.

A first possibility is that, after the inclusion of string-loop
corrections and of possible non-perturbative effects associated
with supersymmetry breaking and with the stabilizitaion of the
dilaton VEV, the Yukawa couplings have no residual
moduli dependence. In this case, the non-vanishing ones will
still obey a superunification condition of the form (\ref{su}),
with $k_i = {\cal O}(1)$. In particular, the top and bottom
Yukawa couplings will fall in the domain of attraction of the
infrared fixed curve discussed in section~2.

A second possibility is that for some Yukawa couplings
there is a residual moduli dependence along some approximately flat
directions. In the following section, we shall assume that this
moduli dependence preserves the unification relation $\hbu = \htauu$,
but allows the top and bottom Yukawa couplings to be treated as
independent variables of the effective theory at the electroweak
scale. Of course, more complicated situations could also arise,
for example that there be only one independent flat direction in
moduli space. In this case, still assuming for simplicity $\hbu =
\htauu$, the allowed range of variation for $\hbu$ and $\htu$ would
be restricted to a certain curve of the $(\htu,\hbu)$ plane, and
this should be taken into account when minimizing the low-energy
effective potential. We shall temporarily disregard this last
possibility in the following section, but we shall come back to it
in the concluding one.

\vspace{1cm}
{\bf 4.}
If, as suggested by the models of ref.~[\ref{lhc}], there are no
quantum corrections to the vacuum energy carrying positive powers
of superheavy scales, the one-loop effective potential of the MSSM
can be written as $V = V_0 + \Delta V$, where
\be
\label{vzero}
V_0 = m_1^2 v_1^2 + m_2^2 v_2^2 + 2 m_3^2 v_1 v_2
+ {g^2+g'^2 \over 8} (v_1^2-v_2^2)^2 + \eta
\, ,
\ee
and
\be
\label{deltav}
\Delta V =
{1 \over 64 \pi ^2} \sum_i (-1)^{2J_i+1} (2J_i+1)
m_i^4 (\log {m_i^2 \over Q^2}-{3 \over 2} )
\, .
\ee
In eq.~(\ref{vzero}), $v_1$ and $v_2$ are the neutral Higgs vacuum
expectation values, and $m_1^2 = m_{H_1}^2 + \mu^2$, $m_2^2 =
m_{H_2}^2 + \mu^2$, $m_3^2 = B \mu$ are mass parameters. The
`cosmological term' $\eta \equiv \hat{\eta} m_{3/2}^4$ takes
into account, as in [\ref{kpz}] but in a slightly different
notation, the contributions to the low-energy effective potential
that do not depend on the MSSM fields. Given a set of boundary
conditions at $M_U$, the mass parameters $m_1^2,m_2^2,m_3^2$ have
an implicit dependence on the top and bottom Yukawa couplings via
their RGEs, as illustrated by the approximate analytical solutions
given in the Appendix. A similar implicit dependence is present
for the parameter $\eta$, whose renormalization-group evolution
was studied in [\ref{kpz}]. In eq.~(\ref{deltav}), $m_i$ and $J_i$
are the tree-level field-dependent mass and the spin for each
particle
$i$ in the MSSM spectrum. Notice that $\Delta V$ has an explicit
dependence on $h_t$ and  $h_b$ only via the top, bottom, stop and
sbottom squared masses,
\be
m_t^2 = h_t^2 v_2^2 \, ,
\;\;\;\;\;
m_b^2 =  h_b^2 v_1^2 \, ,
\ee
\bea
m_{\tilde{t}_{1,2}}^2 & = &
\displaystyle{
h_t^2 v_2^2 + \frac{m_{Q_3}^2 + m_{U_3}^2}{2} + \frac{g^2 + g'^2}{8}
(v_1^2 - v_2^2)} \nonumber \\ & & \nonumber \\
& \pm & \displaystyle{ \sqrt{ \left[  \frac{m_{Q_3}^2 - m_{U_3}^2}{2}
+ \frac{3 g^2 - 5 g'^2}{24} (v_1^2 - v_2^2) \right]^2 +
h_t^2 (A_t v_2 + \mu v_1)^2 }} \, ,
\eea
\bea
m_{\tilde{b}_{1,2}}^2 & = & \displaystyle{
h_b^2 v_1^2 + \frac{m_{Q_3}^2 + m_{D_3}^2}{2} - \frac{g^2 + g'^2}{8}
(v_1^2 - v_2^2)} \nonumber \\ & & \nonumber \\
& \pm & \displaystyle{ \sqrt{ \left[  \frac{m_{Q_3}^2 - m_{D_3}^2}{2}
- \frac{3 g^2 + 2 g'^2}{24} (v_1^2 - v_2^2) \right]^2 +
h_b^2 (A_b v_1 + \mu v_2)^2 }} \, .
\eea

As announced, we shall discuss here the minimization of the MSSM
one-loop effective potential not only with respect to the Higgs
fields, $v_1$ and $v_2$, but also with respect to the top and
bottom Yukawa couplings, $h_t$ and $h_b$, treated as independent
dynamical variables. We would like to stress once more the
importance of the cosmological term $\eta$ in the RG-improved
tree-level potential of eq.~(\ref{vzero}). This term is usually
neglected because it does not depend on $v_1$ and $v_2$; hence it
does not play any significant role in the minimization with respect
to the Higgs fields. In our case, however, this term must be
included, since, given a boundary value $\eta_0 \equiv \eta(M_U)$,
$\eta$ has an implicit dependence on the Yukawa couplings via its
renormalization group evolution: neglecting $\eta$ would create an
artificial dependence of the effective potential on the
renormalization scale $Q$.

With the above comments in mind, we can proceed to the minimization
of the one-loop effective potential with respect to the Higgs
fields and the top and bottom Yukawa couplings, given a set of
boundary conditions $(m_0,m_{1/2}, A_0,B_0,\mu_0 \, ; \eta_0)$.
For convenience, we work with the polar coordinates $\rho$ and
$\theta$ already introduced in section~2, and we proceed in two
separate steps. First, we fix $\theta$ to an arbitrary value, and
we minimize the potential with respect to $v_1$, $v_2$ and $\rho$.
We find that, for any given value of $\theta$, the value of $V$ at
its minimum with respect to $v_1$ and $v_2$ gets smaller and smaller
as $\rho$ increases, until $\rho$ reaches its maximum allowed value,
$\rho_{IR} (\theta)$, corresponding to a point on the effective
infrared fixed line. This result has been tested numerically for
many different values of $\theta$ and of the boundary conditions
on the free parameters, using the full one-loop effective potential
of eqs.~(\ref{vzero}) and (\ref{deltav}). We also verified that in
most cases, with an appropriate choice of the renormalization scale,
$Q^2 \sim m_{\tilde{t}_1} m_{\tilde{t}_2}$, the minimization with
respect to $h_t$ and $h_b$ is dominated by the $V_0$ contribution:
this extends the results obtained in [\ref{grz}] for the usual
minimization with respect to $v_1$ and $v_2$.  The mechanism of
attraction towards the effective infrared fixed line can be
understood semi-analytically in sufficiently simple cases, as we
shall now discuss on an example.

Consider the toy version of the MSSM corresponding to $m_0 =
A_0 = B_0 = \mu_0 = 0$, $v_1=0$, $h_b=0$, but with $m_{1/2}$
and $\eta_0$ both taken as fixed numerical inputs and not as
dynamical variables. In this case, eq.~(\ref{vzero}) simplifies
to
\be
V_0 = m_2^2 v_2^2 + {g^2 + g'^2 \over 8} v_2^4 + \eta \, .
\ee
Assuming $m_2^2 < 0$, as needed for $SU(2) \times U(1)$ breaking,
and minimizing with respect to $v_2$, we find $\left. V_0
\right|_{v_2= \langle v_2^2 \rangle} = - 2 m_2^4 / (g^2 + g'^2)
+ \eta$, and finally
\be
\label{v0min}
\left. {\partial V_0 \over \partial x} \right|_{v_2=\langle v_2
\rangle}
={1 \over 2 \pi} {- 2 m_2^2 \over \alpha_2 + \alpha'}
{\partial m_2^2 \over \partial x} +{\partial \eta \over \partial x}
\, ,
\ee
where $x \equiv \alpha_t / \alpha_t^{IR} = 2 \pi E \alpha_t / (3 F)$.
The Yukawa-coupling dependence of $m_2^2$ can be easily understood
by specializing the general formulae given in the Appendix,
\be
m_2^2 = (C+Ax+Bx^2) m_{1/2}^2 \, ,
\ee
where, at scales $Q$ of the order of the electroweak scale and in
the notation of the Appendix, $C = C_1/4 + C_2 \simeq 0.5$, $A =
- H_2/2 \simeq -5$, $B = H_4^2 / 8 \simeq 2$. We can check
the well-known fact that, in the physical region
$x < 1$, we always have $Ax+Bx^2 < 0$, which leads to $m_2^2
< 0$ for sufficiently large values of $x$. If the $\eta$-dependent
part of eq.~(\ref{v0min}) can be neglected, we can already state
that in the case under consideration $x$ is driven to $x=1$.
However, we know from [\ref{kpz}] that $\eta$ (slowly) increases
for increasing $x$, thus a quantitative comparison of the two terms
in eq.~(\ref{v0min}) is necessary. From the RGE for $\eta$ and
$\alpha_t$, we obtain
\be
{\partial \eta \over \partial x} =
- {1 \over 2 \pi} {(A+Bx)(D+Ax+Bx^2)
\over  G +  J  (1-x)} m_{1/2}^4 \, ,
\ee
where, at scales $Q$ of the order of the electroweak scale and in
the notation of the Appendix, $D = 2C_3 + 2 C_2 + 13 C_1 / 18
\simeq 11.3$, $G = 16 \alpha_3 / 3 + 3 \alpha_2 + 13 \alpha' / 9
- 6 \alpha_t^{IR} \simeq 0.01$, $J = 6 \alpha_t^{IR} \simeq 0.57$.
{}From this one can easily verify that the $\eta$-dependent part of
eq.~(\ref{v0min}) is indeed negligible for all values of $x < 1$.

Having obtained the result that, for any given value of $\theta$,
minimization with respect to $\rho$ invariably leads to $\rho =
\rho_{IR} (\theta)$, we can now restrict our attention to top and
bottom Yukawa couplings constrained along the effective infrared
fixed line, and minimize the effective potential of the MSSM with
respect to the residual angular variable $\theta$ (in addition to
the usual variables $v_1$ and $v_2$). Numerical investigations
show that, depending on the chosen boundary conditions
for the mass parameters, different structures may appear.
A typical situation is illustrated, for a representative parameter
choice, in fig.~3: it corresponds to a trivial minimum for $\theta =
0$, i.e. to vanishing bottom and tau tree-level masses. As we shall
discuss later, this type of structure can be rescued by some
constraint on the moduli space that forbids the boundary
condition $\theta_U=0$. In this case, the low-energy
$\theta$ just relaxes to its minimum allowed value, $\theta_{min}
\ne 0$, and a hierarchy $0 < m_b/m_t \ll m_t$ can emerge even
for values of $\tan \beta$ close to 1. In our numerical
investigations, we were not able to find unconstrained non-trivial
minima for $\theta \ne 0$, corresponding to universal boundary
conditions on the mass parameters and a particle spectrum
compatible with the present experimental data. Establishing whether
realistic solutions of this kind can be obtained or not would
require further investigations. If such solutions do exist, minima
close to $\theta = \pi/4$ would be favoured by the peculiar behaviour
of the function $f(\sin^2 2 \theta)$ and by the focusing effect of
the RGEs. When $\theta \sim \pi/4$, acceptable values for
$m_t$ and $m_b$ can be obtained only for $\tb \sim m_t/m_b$.
This situation can be realized either by selecting a strongly
restricted region of parameter space or by allowing some violation
of universality in the boundary conditions for the mass parameters.

\vspace{1cm}
{\bf 5.}
Motivated by four-dimensional superstring models and their effective
supergravity theories, we examined the possibility of treating the
Yukawa couplings of the MSSM as dynamical variables at the
electroweak scale. In particular, we concentrated on the Yukawa
couplings of the third generation, neglecting the two light
generations and assuming the unification relation $\hbu = \htauu$.
We have found that, treating $(\htu,\hbu)$ as independent variables,
minimization of the one-loop MSSM effective potential attracts
$(h_t,h_b)$ to an effective infrared fixed line. This general feature
allows the elimination of one of the free parameters of the MSSM, and
leads to the generic prediction
\be
{8 \over 9} \alpha_3 \simlt \alpha_t +
\alpha_b \simlt {32 \over 21} \alpha_3 \, ,
\ee
which can be further improved by including $\alpha_{\tau}$,
$\alpha'$,
higher-loop and threshold corrections. In terms of the top and bottom
quark running masses, the above prediction reads
\be
(M_t^{IR})^2 \simlt {m_t^2 \over \sin^2 \beta} +
{m_b^2 \over \cos^2 \beta} \simlt {12 \over 7}
(M_t^{IR})^2 \, ,
\ee
where $M_t^{IR} \simeq (4/3)\sqrt{\alpha_3/(\alpha_2+\alpha')}
m_Z \simeq 195 \gev$. This result can be translated into a relation
involving the pole top and bottom masses by straightforward
inclusion of some finite MSSM one-loop corrections, dominated by
standard QCD effects. The ratio $\alpha_b/\alpha_t$ is also
determined by minimization, but its actual value at the minimum
depends on the free mass parameters of the MSSM $(m_0, m_{1/2},
A_0, B_0, \mu_0 \, ; \eta_0)$. The number of the MSSM free
parameters is further reduced by one, but no generic prediction
can be made in the absence of a theory of the mass parameters.

Our results were obtained under two important assumptions.
First, the overall scale $\msu$ of the MSSM mass parameters,
proportional to the gravitino mass, was not taken as a
dynamical variable but as a fixed numerical input. Second,
the two Yukawa couplings $\htu$ and $\hbu=\htauu$ were
considered as independent variables in the minimization.
We would like to conclude our paper by commenting on the
effects of relaxing each of these two assumptions.

When, as in [\ref{kpz}], $\msu$ is also considered as a dynamical
variable, one obtains an additional constraint on the MSSM mass
parameters, coming from the minimization condition with respect
to the gravitino mass, which sets the overall MSSM mass scale,
\be
\label{master}
m_{3/2}^2 {\partial V_1 \over \partial m_{3/2}^2} = 2 V_1 +
{{\rm Str \,} {\cal M}^4 \over 64 \pi^2} = 0 \, .
\ee
It was shown in [\ref{kpz}] that the above equation allows
for the dynamical generation of the desired $\msu / \mpl$
hierarchy in a large region of the parameter space. The main
quantitative result on the dynamical determination of the Yukawa
couplings, i.e. the generic attraction of $\alpha_t+\alpha_b$
towards the effective infrared fixed line, remains the same.
However, the determination of the ratio $\alpha_b / \alpha_t$
as a function of the boundary conditions for the residual free
mass parameters would require a separate analysis, since the
minimization condition with respect to $m_{3/2}$, eq.~(\ref{master}),
induces a non-trivial dependence of the overall mass scale of the
effective potential on the angular variable $\theta$ parametrizing
the effective infrared fixed line: this might allow for an easier
generation of phenomenologically acceptable non-trivial minima at
$\theta \ne 0$.

Another possibility is that the top and bottom Yukawa couplings
at the unification scale, $\hbu$ and $\htu$, are not independent
but constrained by some functional relation, corresponding to a
curve in the $(\htu,\hbu)$ plane, such as those shown in fig.~1c.
Such a possibility can occur if the moduli dependences of $\hbu$
and $\htu$ are correlated, and correspond to a single independent
flat direction in moduli space: in this case also the number of
independent minimization conditions has to be restricted accordingly.
However, the generic phenomenon of attraction towards the effective
infrared fixed line will persist also in this case, as long as the
constraint on $(\htu,\hbu)$ allows for some points with sufficiently
large $\rho_U$ to fall in its domain of attraction. If the constraint
at the unification scale allows for all possible values of $\theta_U$
within the domain of attraction of the effective infrared fixed line,
then such a curve is mapped into the entire infrared fixed line, and,
as far as low-energy Yukawa couplings are concerned, minimization
under the constraint gives exactly the same result as unconstrained
minimization. It might well be, however, that either the range of
variation of $\theta_U$ is restricted, or a sufficently large value
of $\rho_U$ is allowed only for certain values of $\theta_U$, as is
the case for the dot-dashed and solid curves in fig.~1c,
respectively.
In this case the minima of the low-energy potential will still lie
on the infrared fixed line, but minimization with respect to $\theta$
must take into account the bounds set by the constraint at the
unification scale, as apparent from the corresponding curves in
fig.~1d. The only case in which the constrained minimum does not lie
along the effective infrared fixed line corresponds to a curve in the
$(\htu,\hbu)$ plane that does not allow for sufficiently large values
of $\rho_U$: we regard this last situation, exemplified by the dashed
lines in figs.~1c and~1d, as extremely unlikely, given the fact that
the
tree-level string Yukawa couplings are typically of the order of the
unified gauge coupling, which falls already in the domain of
attraction of the effective infrared fixed line. Several possible
constraints on the Yukawa couplings at the unification scale were
recently conjectured in [\ref{binetruy}], but without dwelling into
a possible string origin. We have argued here that the detailed form
of these constraints may or may not be relevant for the determination
of the low-energy Yukawa couplings. The present understanding of the
moduli space of four-dimensional superstring models, with its
generalized
duality symmetries, indeed suggests the possible existence of such a
constraint, but does not allow to single out a specific form for it
in a model-independent way.

\section*{Acknowledgements}
We thank M.~Carena and C.E.M.~Wagner for many useful conversations.

\newpage
\section*{Appendix}
We present here some approximate analytical solutions to the RGE for
the MSSM mass parameters, which include $h_t$ and $h_b$ effects for
any values of the latter, but still neglect $h_{\tau}$ effects. We
assume for simplicity universal boundary conditions at the
unification
scale:
\be
M_3 (M_U) = M_2 (M_U) = M_1 (M_U) \equiv m_{1/2},
\ee
$$
\tilde{m}_{Q_a}^2 (M_U)
=\tilde{m}_{U^c_a}^2 (M_U)
=\tilde{m}_{D^c_a}^2 (M_U)
=\tilde{m}_{L_a}^2 (M_U)
$$
\be
=\tilde{m}_{E^c_a}^2 (M_U)
=m_{H_1}^2 (M_U)
=m_{H_2}^2 (M_U)
\equiv m_0^2 \, ,
\ee
\be
A^U (M_U) = A^D (M_U) = A^E (M_U) \equiv A_0
\, , \;\;\;
B (M_U) \equiv B_0
\, , \;\;\;
\mu(M_U) \equiv \mu_0 \, .
\ee
To parametrize the dependence of the relevant mass parameters
on the top and bottom Yukawa couplings, we introduce the auxiliary
variables
\be
\label{xy}
x \equiv \left( {h_t \over \sqrt{{8 \pi^2 E }\over { 3 F } } }
\right)^2
\, ,
\;\;\;\;\;
y \equiv \left( {h_b \over \sqrt{{8 \pi^2 E }\over { 3 F } } }
\right)^2
\, .
\ee
On the infrared curve of eq.~(\ref{hthb0}), we can write $x =
f(\sin^2
2\theta) \cos^2 \theta$ and $y = f (\sin^2 2 \theta) \sin^2 \theta$.
To include the most important $\alpha'$ effects, in such a way that
our approximate formulae are optimized for $x \simgt y$, we define
\be
E \equiv Z_3^{ {16 \over 9}} Z_2^{- {3}} Z_1^{-{13 \over 99}}\, ,
\;\;\;\;
F \equiv \int_0^t E(t') dt' \, ,
\ee
where ($i=1,2,3$)
\be
Z_i \equiv \left( 1 + {b_i \over 4 \pi} t \right)^{-1} \, ,
\ee
and
\be
b_3=-3 \, , \;\;\;\; b_2=1 \, , \;\;\;\; b_1= {33 \over 5} \,.
\ee
The low-energy mass parameters with a non-trivial dependence on
$x$ and $y$ can be written as
\be
m_{Q_3}^2=m_0^2+ ({1 \over 36}C_1 +C_2 + C_3) m_{1/2}^2-{1 \over 3}
\left( \Delta m_1^2 + \Delta m_2^2 \right) \, ,
\ee
\be
m_{U_3}^2=m_0^2+({4 \over 9}C_1 + C_3)m_{1/2}^2-{2 \over 3} \Delta
m_2^2 \, ,
\ee
\be
m_{D_3}^2=m_0^2+({1 \over 9} C_1 + C_3)m_{1/2}^2-{2 \over 3} \Delta
m_1^2 \, ,
\ee
\be
m_{H_1}^2=m_0^2+({1 \over 4} C_1 +C_2)m_{1/2}^2-\Delta m_1^2 \, ,
\ee
\be
m_{H_2}^2=m_0^2+({1 \over 4} C_1 +C_2)m_{1/2}^2-\Delta m_2^2 \, ,
\ee
\be
\mu^2= \mu_0^2 \left( {\alpha_t \over \atu} {\alpha_b \over \abu}
\right)^{3/7} Z_3^{-32/21} Z_2^{-3/7} Z_1^{-1/231} \, ,
\ee
\be
B=B_0 -{1 \over 2} A_0 (x+y) +m_{1/2}[H_9-{1 \over 4} H_4 (x+y)] \, ,
\ee
\be
A_t=A_0(1-x-{1 \over 6}y)+m_{1/2}[H_8-{1 \over 2}H_4(x+{1 \over 6}y)]
\, ,
\ee
\be
A_b=A_0(1-{1 \over 6}x-y)+m_{1/2}[\tilde H_8-{1 \over 2}H_4({1 \over
6}x+y)] \, ,
\ee
\be
A_{\tau}=A_0 (1-{1 \over 2}y) + m_{1/2}( H_{10}-{1 \over 4} H_4 y) \,
,
\ee
where
\be
C_1 \equiv \frac{2}{11} (1-Z_1^2) \, ,
\;\;\;\;\;
C_2 \equiv \frac{3}{2} (1-Z_2^2) \, ,
\;\;\;\;\;
C_3 \equiv - \frac{8}{9} (1-Z_3^2) \, .
\ee
The only two independent quantities entering the solutions for the
soft scalar masses are
\be
\Delta m_1^2={3 \over 2}m_0^2 y
+ {1 \over 2} A_0 y [1-a(x,y)y] ( H_4 m_{1/2} + A_0 )
+ {1 \over 2} m_{1/2}^2 y [ \tilde H_2 -{1 \over 4} \; a(x,y) H_4^2
y]
\, ,
\ee
\be
\Delta m_2^2={3 \over 2}m_0^2 x
+ {1 \over 2} A_0 x [1-a(x,y)x] ( H_4 m_{1/2} + A_0 )
+ {1 \over 2} m_{1/2}^2 x [ H_2 -{1 \over 4} \; a(x,y) H_4^2 x]
\, ,
\ee
where
\be
a(x,y)={7 f [ {4xy \over (x+y)^2} ] + 23 \over 30}
\ee
is a suitable interpolating function, such that $a(x,0)=a(0,y)=1$,
$a(x,y=x)=7/6$, and
\be
H_2 \equiv \frac{E}{F} t H_8 \, ,
\;\;\;\;\;
\tilde H_2 \equiv \frac{E}{F} t \tilde H_8 \, ,
\;\;\;\;\;
H_4 \equiv 2 \left( t \frac{E}{F} - 1 \right) \, ,
\ee
\be
H_8 \equiv \frac{\alpha_U}{4 \pi} t \left( \frac{16}{3} Z_3 + 3 Z_2 +
\frac{13}{15} Z_1 \right)
\, , \;\;\;\;
H_9 \equiv {\alpha_U \over {4 \pi}} t (3 Z_2+{3 \over 5} Z_1) \, ,
\ee
\be
\tilde H_8 \equiv \frac{\alpha_U}{4 \pi} t \left( \frac{16}{3} Z_3 +
3 Z_2 + \frac{7}{15} Z_1 \right) \, , \;\;\;\;
H_{10} \equiv \frac{\alpha_U}{4 \pi} t \left( 3 Z_2 + \frac{9}{5} Z_1
\right) \, .
\ee

The above formulae have been tested numerically by comparing them
with the exact numerical solutions of the one-loop RGE. If one
compares with exact numerical solutions neglecting $\atau$ effects,
our approximate results are correct with less than 3\% error. When
$\atau$ effects are included in the comparison, the error of our
formulae grows up to a maximum of 10\%.

\newpage
\section*{References}
\begin{enumerate}
\item
\label{kpz}
C.~Kounnas, I.~Pavel and F.~Zwirner, Phys. Lett. B335 (1994) 403.
\item
\label{similar}
Y.~Nambu, Univ. of Chicago preprint EFI 92-37;
\\
V.I.~Zakharov, in `Properties of SUSY particles' (L.~Cifarelli
and V.A.~Khoze eds., World Scientific, Singapore, 1993), p.~65.
\item
\label{binetruy}
P.~Bin\'etruy and E.~Dudas, Phys. Lett. B338 (1994) 23 and preprint
LPTHE Orsay 94/73, SPhT Saclay T94/145, hep-ph/9411413.
\item
\label{noscale}
E.~Cremmer, S.~Ferrara, C.~Kounnas and D.V.~Nanopoulos,
Phys. Lett. B133 (1983) 61;
\\
J.~Ellis, A.B.~Lahanas, D.V.~Nanopoulos and K.~Tamvakis, Phys. Lett.
B134 (1984) 429;
\\
J.~Ellis, C.~Kounnas and D.V.~Nanopoulos,
Nucl. Phys. B241 (1984) 406 and B247 (1984) 373;
\\
R.~Barbieri, S.~Ferrara and E.~Cremmer, Phys. Lett. B163 (1985) 143.
\item
\label{lhc}
S.~Ferrara, C.~Kounnas, M.~Porrati and F.~Zwirner,
Phys. Lett. B194 (1987) 366 and Nucl. Phys. B318 (1989) 75;
\\
S.~Ferrara, C.~Kounnas and F.~Zwirner, Nucl. Phys. B429 (1994)  589;
\\
A.~Brignole and F.~Zwirner, Phys. Lett.~B342 (1995) 117.
\item
\label{btauns}
M.~Chanowitz, J.~Ellis and M.K.~Gaillard, Nucl. Phys. B128 (1977)
506;
\\
A.~Buras, J.~Ellis, M.K.~Gaillard and D.V.~Nanopoulos, Nucl. Phys.
B135 (1978) 66.
\item
\label{btaususy}
M.B.~Einhorn and D.R.T.~Jones, Nucl. Phys. B196 (1982) 475;
\\
K.~Inoue, A.~Kakuto, H.~Komatsu and S.~Takeshita, Progr.
Theor. Phys. 67 (1982) 1889.
\item
\label{hill}
B. Pendleton and G.G. Ross, Phys. Lett. B98 (1981) 291;
\\
C. Hill, Phys. Rev. D24 (1981) 691.
\item
\label{apw}
K.~Inoue, A.~Kakuto, H.~Komatsu and S.~Takeshita, Progr.
Theor. Phys. 67 (1982) 1889;
\\
L.~Alvarez-Gaum\'e, J.~Polchinski and M.B.~Wise, Nucl. Phys. B221
(1983) 495;
\\
J. Bagger, S. Dimopoulos and E. Mass\'o, Phys. Rev. Lett. 55 (1985)
920.
\item
\label{rest}
H.~Arason, D.J.~Casta\~no, B.~Keszthelyi, S.~Mikaelian, E.J.~Piard,
P.~Ramond and B.D.~Wright, Phys. Rev. Lett. 67 (1991) 2933;
\\
M.~Carena, T.E.~Clark, C.E.M.~Wagner, W.A.~Bardeen and K.~Sasaki,
Nucl. Phys. B369 (1992) 33;
\\
S.~Kelley, J.L.~Lopez and D.V.~Nanopoulos, Phys. Lett. B278 (1992)
140;
\\
H.E.~Haber and F.~Zwirner, unpublished, as quoted in H.E.~Haber,
`Properties of SUSY Particles', (L.~Cifarelli and V.A.~Khoze eds.,
World Scientific, Singapore, 1993), p.~321;
\\
V.~Barger, M.S.~Berger, P.~Ohmann and R.J.N.~Phillips, Phys. Lett.
B314 (1993) 351;
\\
M.~Carena, S.~Pokorski and C.E.M.~Wagner, Nucl. Phys. B406 (1993) 59;
\\
W.A.~Bardeen, M.~Carena, S.~Pokorski and C.E.M.~Wagner, Phys. Lett.
B320 (1994) 110;
\\
P.~Langacker and N.~Polonsky, Phys. Rev. D49 (1994) 1454.
\item
\label{cw}
M.~Carena, M.~Olechowski, S.~Pokorski and C.E.M.~Wagner,
Nucl. Phys. B419 (1994) 213.
\item
\label{largetb}
M.~Carena, M.~Olechowski, S.~Pokorski and C.E.M.~Wagner,
Nucl. Phys. B426 (1994) 269;
\\
L.J.~Hall, R.~Rattazzi and U.~Sarid, Phys. Rev. D50 (1994) 7048;
\\
R.~Rattazzi, U.~Sarid and L.J.~Hall, Stanford Univ. preprint
SU-ITP-94-15, hep-ph/9405313;
\\
M.~Carena and C.E.M.~Wagner, preprint CERN-TH.7321/94.
\item
\label{th7320}
M.~Carena and C.E.M.~Wagner, preprint CERN-TH.7320/94.
\item
\label{fl}
E.G.~Floratos and G.K.~Leontaris, Phys. Lett. B336 (1994) 194.
\item
\label{rge}
R. Barbieri, S. Ferrara, L. Maiani, F. Palumbo and C.A. Savoy, Phys.
Lett. B115 (1982) 212.
\item
\label{twoloop}
J.E.~Bj\"orkman and D.R.T.~Jones, Nucl. Phys. B259 (1985) 533;
\\
V.~Barger, M.S.~Berger and P.~Ohmann, Phys. Rev. D47 (1993) 1093.
\item
\label{numacc}
M.~Carena et al., as in ref.~[\ref{rest}].
\item
\label{massrge}
K.~Inoue et al., as in ref.~[\ref{apw}];
\\
L.~Alvarez-Gaum\'e et al., as in ref.~[\ref{apw}];
\\
J.-P.~Derendinger and C.A.~Savoy, Nucl. Phys. B237 (1984) 307;
\\
B.~Gato, J.~Leon, J.~P\'erez-Mercader and M.~Quir\'os,
Nucl. Phys. B253 (1985) 285;
\\
N.K.~Falck, Z.~Phys. C30 (1986) 247.
\item
\label{extonly}
L.E.~Ib\'a\~nez and C.~Lopez, Nucl. Phys. B233 (1984) 511;
\\
C.~Kounnas, A.B.~Lahanas, D.V.~Nanopoulos and M.~Quir\'os, Nucl.
Phys.
B236 (1984) 438;
\\
L.E.~Ib\'a\~nez, C.~Lopez and C.~Mu\~noz, Nucl. Phys. B256 (1985)
218;
\\
A.~Bouquet, J.~Kaplan and C.A.~Savoy, Nucl. Phys. B262 (1985) 299.
\item
\label{fds}
P.~Candelas, G.~Horowitz, A.~Strominger and E.~Witten, Nucl. Phys.
B258 (1985) 46;
\\
L.~Dixon, J.~Harvey, C.~Vafa and E.~Witten, Nucl. Phys. B261 (1985)
678;
\\
K.S.~Narain, Phys. Lett. B169 (1986) 41;
\\
K.S.~Narain, M.H.~Sarmadi and E.~Witten, Nucl. Phys. B279 (1987) 369;
\\
W.~Lerche, D.~L\"ust and A.N.~Schellekens, Nucl. Phys. B287 (1987)
477;
\\
H.~Kawai, D.C.~Lewellen and S.-H.H. Tye, Nucl. Phys. B288 (1987) 1;
\\
I.~Antoniadis, C.~Bachas and C.~Kounnas, Nucl. Phys. B289 (1987) 87.
\\
K.S.~Narain, M.H.~Sarmadi and C.~Vafa, Nucl. Phys. B288 (1987) 551.
\\
For a review and further references see, e.g.:
\\
B.~Schellekens, ed., {`Superstring construction'} (North-Holland,
Amsterdam, 1989).
\item
\label{yukstr}
E.~Witten, Phys. Lett. B155 (1985) 151;
\\
A.~Strominger, Phys. Rev. Lett. 55 (1985) 2547;
\\
A.~Strominger and E.~Witten, Commun. Math. Phys. 101 (1985) 341;
\\
S.~Ferrara, C.~Kounnas and M.~Porrati, Phys. Lett. B181 (1986) 263.
\item
\label{thr}
V.S.~Kaplunovsky, Nucl. Phys. B307 (1988) 145;
\\
L.J.~Dixon, V.S.~Kaplunovsky and J.~Louis, Nucl. Phys. B355 (1991)
649;
\\
J.-P.~Derendinger, S.~Ferrara, C.~Kounnas and F.~Zwirner, Nucl. Phys.
B372 (1992) 145 and Phys. Lett. B271 (1991) 307;
\\
G.~Lopez Cardoso and B.A.~Ovrut, Nucl. Phys. B369 (1992) 351;
\\
I.~Antoniadis, K.S.~Narain and T.~Taylor, Phys. Lett. B276 (1991) 37;
\\
I.~Antoniadis, E.~Gava, K.S.~Narain and T.R.~Taylor, Nucl. Phys. B407
(1993) 706;
\\
E.~Kiritsis and C.~Kounnas, preprint CERN-TH.7472/94, hep-th/9501020.
\item
\label{grz}
G.~Gamberini, G.~Ridolfi and F.~Zwirner, Nucl. Phys. B331 (1990) 331.
\end{enumerate}
\vfill{
\section*{Figure captions}
\begin{itemize}
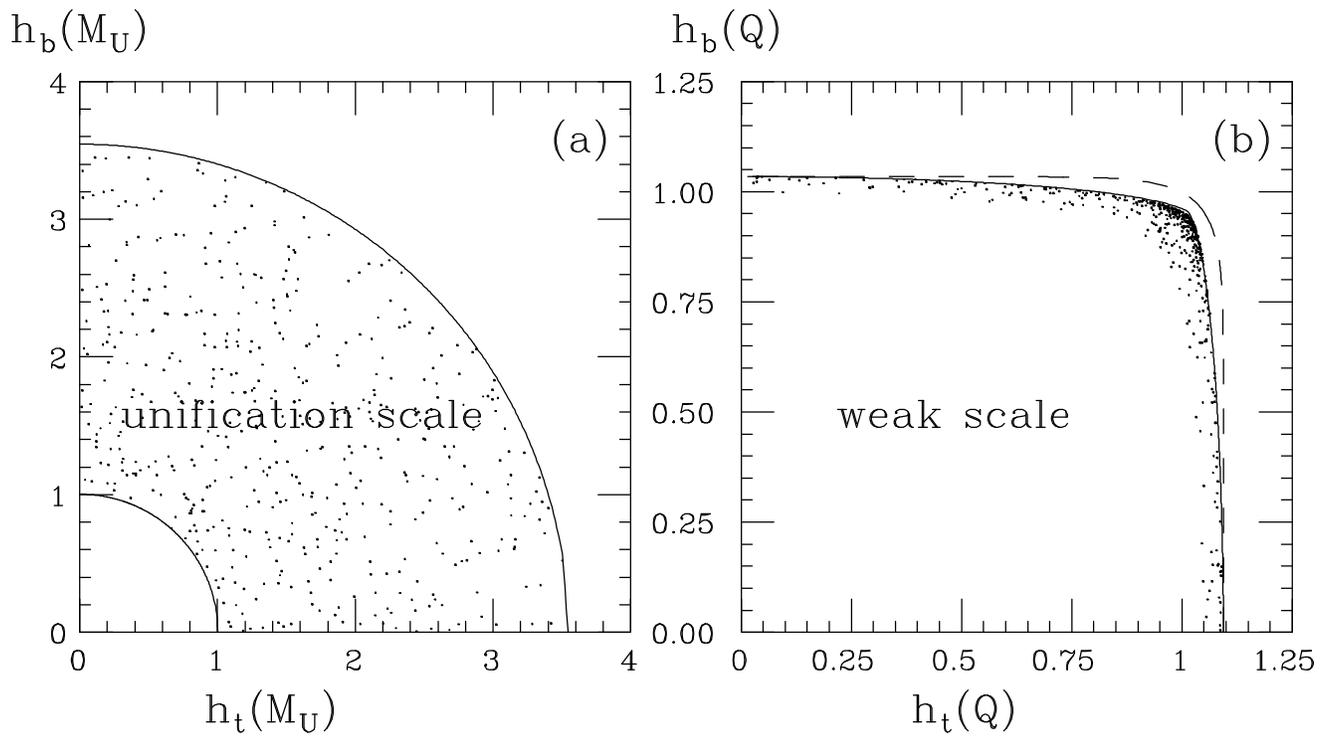
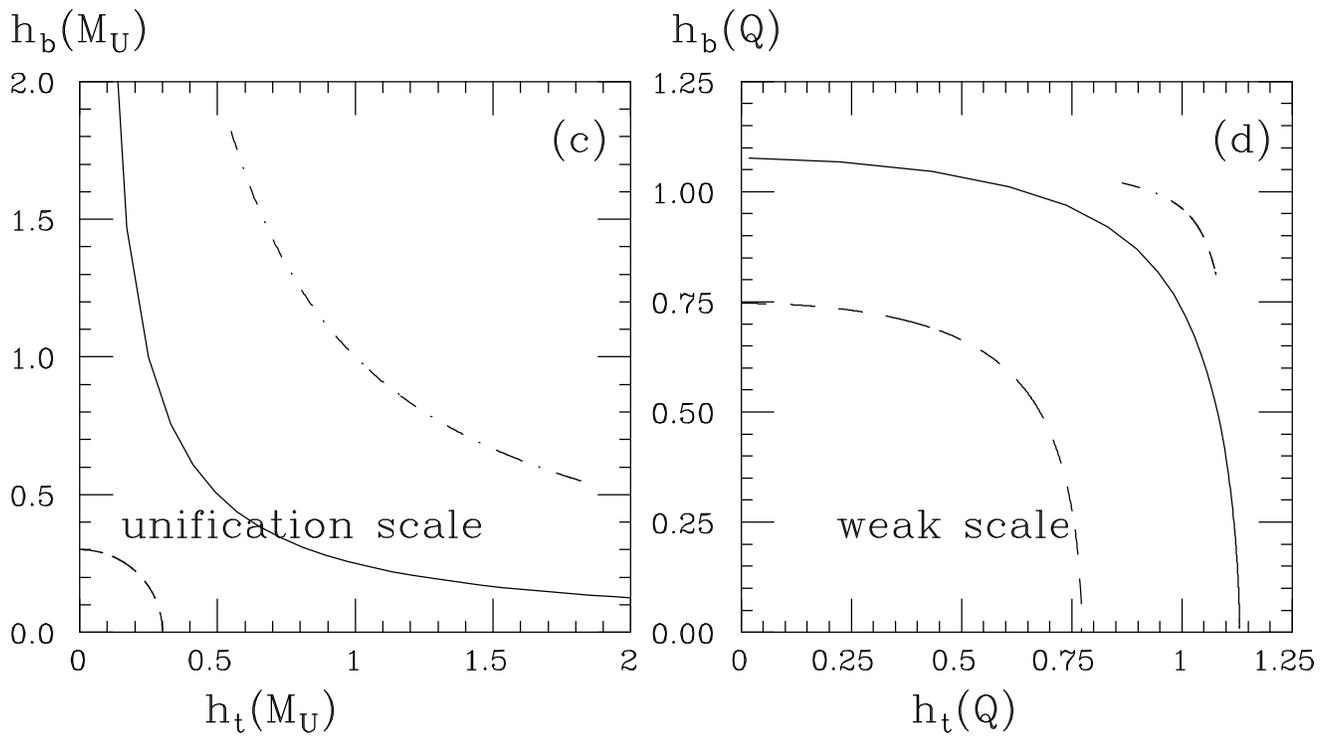
\item[Fig.1:]
Mapping of the $(\htu,\hbu)$ plane into the $(h_t,h_b)$ plane, for
$Q=200 \gev$, $\hbu=\htauu$, $M_{\rm U} = 2 \times 10^{16} \gev$,
$\alpha_U = 1/25$. In (b), the dots correspond to the exact numerical
solutions of the one-loop RGE of eqs.~(\ref{rgeyuk}), for the
boundary conditions given in (a); the solid line corresponds to the
approximate analytical solution of eq.~(\ref{hthb}), and the dashed
line to the approximate solution of eq.~(\ref{power}). In (c) and
(d), we show how some possible constraints, corresponding to curves
in the $(\htu,\hbu)$ plane, are mapped into corresponding curves in
the $(h_t,h_b)$ plane.
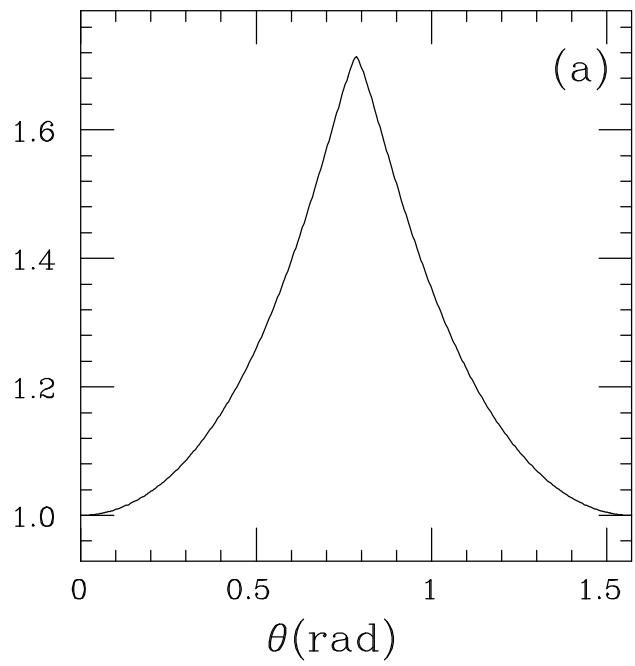
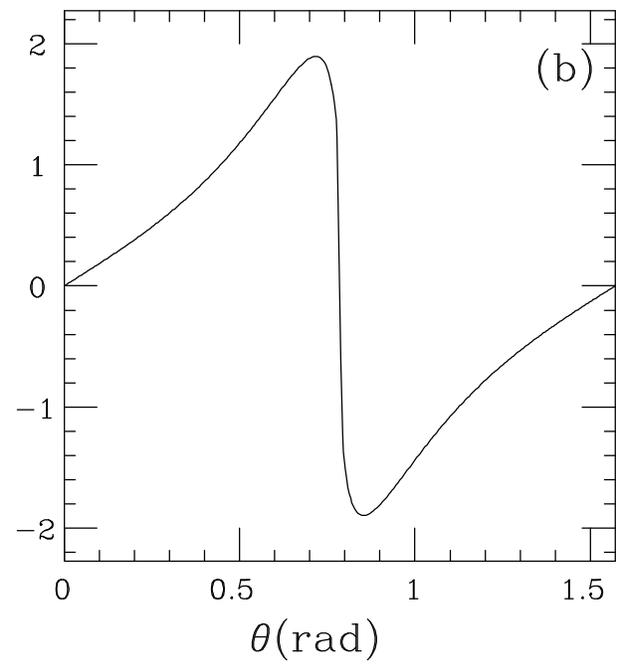
\item[Fig.2:]
The function $f(\sin^2 2 \theta)$ and its derivative $df(\sin^2 2
\theta) /d \theta$.
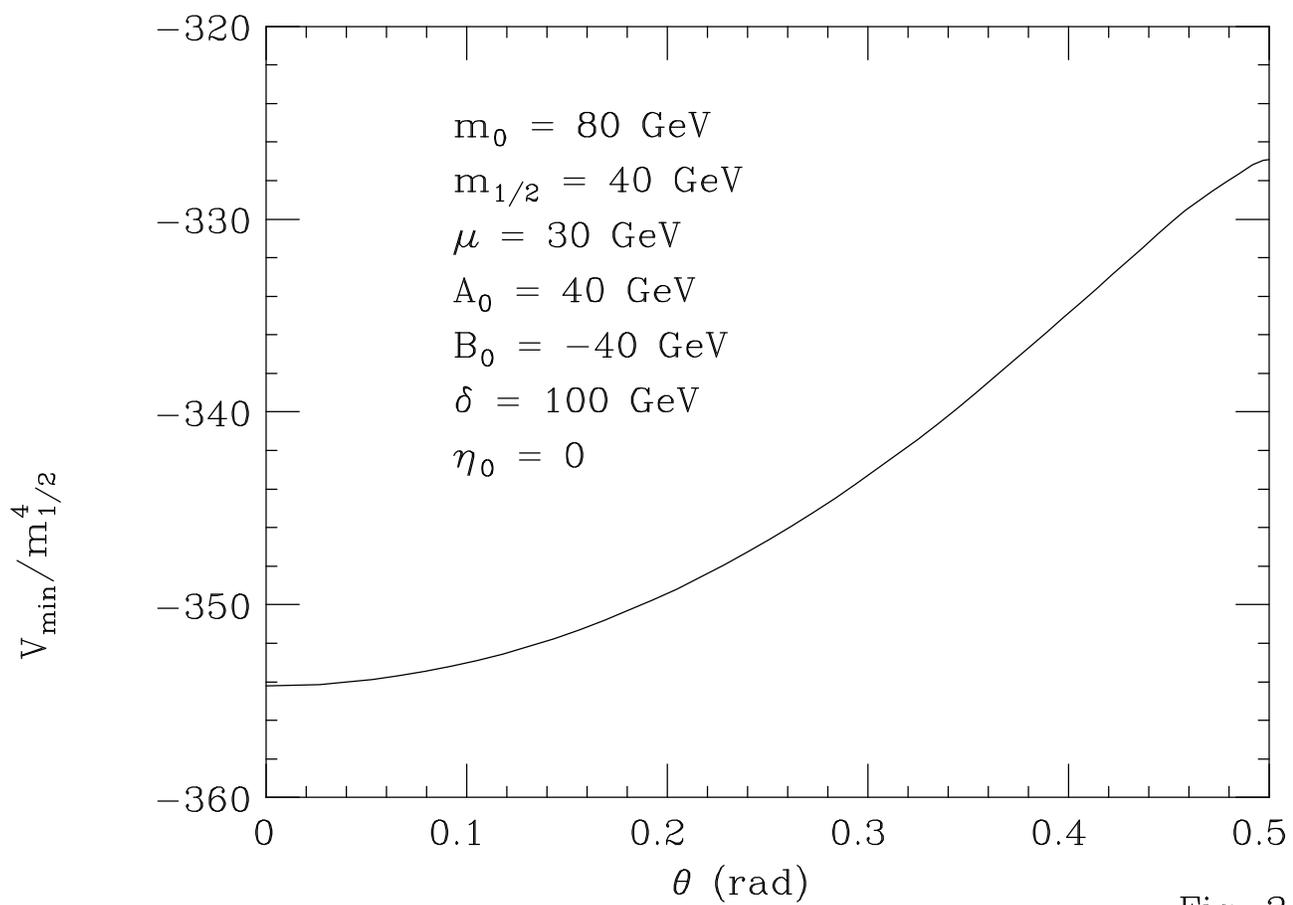
\item[Fig.3:]
$V(\theta)$ for a representative choice of the boundary conditions
$(m_0, m_{1/2}, A_0, B_0, \mu_0)$ and for $\eta_0=0$. For
convenience, a non-universal contribution $\delta$ to the boundary
condition on $m_{H_1}$ has been allowed: $m_{H_1}^2(M_{\rm U}) =
m_0^2 + \delta^2$.
\end{itemize}
}
\end{document}